\documentstyle[nato]{crckapb}
\newcommand{\half}{{\scriptstyle{{1\over 2}}}}
\newcommand{\diag}{\mbox{diag}}
\def\plo{{{\cal P}_\infty^0}}
\def\pl{{{\cal P}_\infty}}
\def\beqa{\begin{eqnarray}}
\def\eeqa{\end{eqnarray}}
\def\beq{\begin{equation}}
\def\eeq{\end{equation}}
\def\hphm{\hphantom{-}}
\def\bea{\begin{array}}
\def\eea{\end{array}}
\def\myre{{\rm Re}}
\def\cA{{\cal{A}}}
\def\cO{{\cal{O}}}
\def\Tr{{\rm Tr}}
\def\tr{{\rm tr}}

\begin{opening}
\title{Chiral zero-mode for abelian BPS dipoles
\vskip-3cm\hfill{\rm INLO-PUB-01/02}\vskip1.4cm                
}
\author{Pierre van Baal}
\institute{Instituut-Lorentz for Theoretical Physics\\
           University of Leiden, P.O.Box 9506\\
           NL-2300 RA Leiden, The Netherlands}
\end{opening}
\begin{document}

\begin{abstract}
We present\footnote{Talk given at the NATO workshop on         
``Confinement, Topology, and other Non-Perturbative Aspects    
of QCD", Star\'a Lesn\'a, Slovakia, January 21-27, 2002.}      
an exact normalisable zero-energy chiral fermion solution for abelian BPS 
dipoles. For a single dipole, this solution is contained within the high 
temperature limit of the SU(2) caloron with non-trivial holonomy.  
\end{abstract}

\section{The Dirac Monopole}

A convenient representation for the Dirac monopole~\cite{Dirac} is given by
\beq
\vec A=g\hat n\wedge\vec\nabla\log w(\vec x),\quad w(\vec x)=
|\vec x|+\vec x\cdot\hat n,
\eeq
where $w(\vec x)$ is positive, but vanishes along the Dirac string pointing 
along $-\hat n$, as seen from the monopole. The magnetic field is 
\beq
\vec B=\vec\nabla\wedge\vec A=-g\vec\nabla\left(\hat n\cdot\vec\nabla\log w(
\vec x)\right)+g\hat n\Delta\log w(\vec x)=\vec B_{\rm reg}+\vec B_{\rm str},
\eeq
and using that $\hat n \cdot\vec\nabla\log w(\vec x)=1/|\vec x|$ (the 
derivative {\em along} the direction of the Dirac string) is {\em independent} 
of $\hat n$, the first term gives rise to the radial magnetic field associated 
with a magnetic point charge, whereas the second term represents the Dirac 
string, which takes care of the return flux. This follows from the fact that 
$\log w(\vec x)$ is harmonic, except where $w(\vec x)$ vanishes. To be
specific, choosing for convenience $\hat n=\hat e_3=(0,0,1)$, we find $\vec B=
g\vec x/|\vec x|^3+4\pi g\hat e_3\delta(x)\delta(y)\theta(-z)$ and $\vec\nabla
\cdot\vec B_{\rm reg}=-\vec\nabla\cdot\vec B_{\rm str}=4\pi g\delta_3(\vec x)$.
It gives the appropriate magnetic point charge for $\vec B_{\rm reg}$, but 
when including the return flux $\vec\nabla\cdot\vec B=0$, as it should.

The function $\log w(\vec x)$ can be viewed as a potential, although from 
the point of view of the Maxwell equations it is more natural to consider 
$A_0\equiv-g\hat n\cdot\nabla\log w(\vec x)$ as such. Not only $\vec B_{
\rm reg}=\vec\nabla A_0$, but also as the time component of the Euclidean 
vector potential this choice of $A_0$ gives rise to a self-dual configuration, 
with $\vec E=\vec\nabla A_0=\vec B_{\rm reg}$. The usefulness of $w(\vec x)$ 
becomes clear when one considers the massless (Euclidean) Dirac equation in 
such a background. As usual we split this into positive and negative 
chirality Weyl equations, 
\beq
H_+=\bar D=D^\dagger=-\bar\sigma^\mu D_\mu,\quad H_-=D=\sigma^\mu D_\mu,
\eeq
where $D_\mu=\partial_\mu+ie A_\mu$ is the covariant derivative and 
$\sigma_j=i\tau_j$, whereas $\sigma_0$ is the $2\times 2$ identity matrix.
For $\hat n=\hat e_3$ a solution of $\bar D\Psi=0$ is given by
\beq
\Psi(\vec x)=\sqrt{w(\vec x)}\pmatrix{-\partial_1+i\partial_2\cr
\partial_3\cr}\log w(\vec x).\label{eq:azm}
\eeq
If so desired a (spin-)rotation allows one to obtain the solution for arbitrary
$\hat n$, but to keep things simple we stick to $\hat n=\hat e_3$, such that
\beq
-iH_+=\pmatrix{\partial_3+i\partial_0&\hphm\partial_1-i\partial_2\cr
               \partial_1+i\partial_2&-\partial_3+i\partial_0\cr}
   -eg\pmatrix{-\partial_3           &-\partial_1+i\partial_2\cr
                 \partial_1+i\partial_2&-\partial_3\cr}\log w(\vec x).
\eeq
Using the Dirac quantisation condition $eg=\half$, one easily verifies
\beq
-iH_+\Psi(\vec x)=\sqrt{w(\vec x)}\pmatrix{0\cr-\Delta\log w(\vec x)\cr}.
\eeq
Since $\sqrt{w}$ vanishes along the Dirac string, we find that $\sqrt{w}
\Delta\log w=0$ (as a distribution), hence $\bar D\Psi=0$. Likewise $|\Psi|^2=
w\vec\nabla\log w\cdot\vec\nabla\log w=\Delta w-w\Delta\log w$. Thus $|\Psi|^2
=\Delta w=2/|\vec x|$ has an integrable singularity at the origin, and the 
Dirac string is {\em invisible}, as it should. Nevertheless, $\Psi$ does not 
decay sufficiently fast to be normalisable. Note that the zero-mode is time 
independent. Putting $\partial_0$ to zero, $H$ is precisely the Dirac 
Hamiltonian, with $A_0$ playing the role of a Higgs field. A non-zero 
asymptotic value of $A_0$ would lead to a mass scale and exponentially 
decaying wave functions\footnote{A constant $A_0$ in the Euclidean Weyl 
equation can also be identified with a non-zero chemical potential.}. 

It should not come as a surprise that existence of zero-energy solutions is 
sensitive to the sign of the electron charge (relative to $g$). With 
$eg=-\half$ we have
\beq
-iH_+\left(\frac{1}{\sqrt{w(\vec x)}}\pmatrix{\partial_3\cr\partial_1+
i\partial_2\cr}\log w(\vec x)\right)=\frac{1}{\sqrt{w(\vec x)}}\pmatrix{
\Delta\log w(\vec x)\cr0\cr}.\label{eq:psit}
\eeq
However, here the singularity of the Dirac string is no longer nullified but
enhanced. Nevertheless, it can be turned into a proper zero-energy solution, 
identical to $\Psi^\dagger\sigma_2$, by replacing $w(\vec x)$ with 
$1/w(\vec x)$, but this has the same effect as changing $g$ to $-g$, 
explaining why the new zero-energy solution is the charge conjugate of $\Psi$ 
in Eq.~(\ref{eq:azm}). Negative chirality zero-energy solutions cannot appear
because the self-duality of $A_\mu$ implies that $\bar D D=-D_\mu^2$. 
Therefore such a solution would satisfy $D_\mu\Psi=0$, which is ruled out.

It is well known that the 't Hooft-Polyakov monopole~\cite{THPo} allows for 
a normalisable chiral zero-energy solution of the Dirac equation~\cite{JaRe}. 
The size of the core of these non-abelian monopoles is determined by the mass 
scale set by the asymptotic value of the Higgs field. When the core size 
shrinks to zero, so does the support for the zero-mode. Adding to the Higgs 
field $\Phi= \half\Phi^a\sigma_a$ a constant element in U(1), $\Phi_z=\Phi-
2\pi i z$, as it appears in Nahm's work~\cite{MonN}, the zero-mode remains 
normalisable for a finite range of $z$ determined by the Callias index 
theorem~\cite{Call}. Our solution corresponds to $z$ at the boundary of 
this range, where the zero-mode fails to be normalisable. This boundary 
value of $z$ is defined by $\det(\Phi_z)=0$, and the abelian field $A_0$ 
given above corresponds to the isospin component of $\Phi_z$ responsible 
for this vanishing eigenvalue. It does imply the support of the zero-mode 
is no longer confined to the non-abelian core.

There has been another context in which solutions to the Dirac equation 
in the background of a monopole have appeared in the past, namely that 
of monopole-induced proton decay (the Callan-Rubakov effect~\cite{CaRu}). 
Boundary conditions~\cite{Gold} for the fermions are imposed at the core 
of the monopole to describe the scattering states in the limit where the 
size of the monopole core can be neglected, so as to properly reflect the 
breaking of B-L, compatible with the chiral anomaly. There it is assumed, as 
for the Jackiw-Rebbi zero-energy solution, that the Higgs field approaches a 
non-zero constant at infinity, which through the Yukawa coupling gives a mass 
to the fermions\footnote{For a non-vanishing Higgs mass, this together with
the Dirac monopole field, is all that is left when neglecting the core of the 
monopole. In the Bogomol'ny limit considered here, identifying the Higgs field 
with $A_0$ (self-duality implied by the BPS equations~\cite{BPS}), the long 
range component of the Higgs field modifies the Dirac equation.}. In our case 
the asymptotic value of the Higgs field ($A_0$) vanishes and as we will see, 
the limit of zero monopole core size can be taken without any approximation, 
but at the expense of the zero-energy state being non-normalisable.

\section{The Abelian BPS Dipole}

\begin{figure}[htb]
\vspace{4.3cm}
\includegraphics{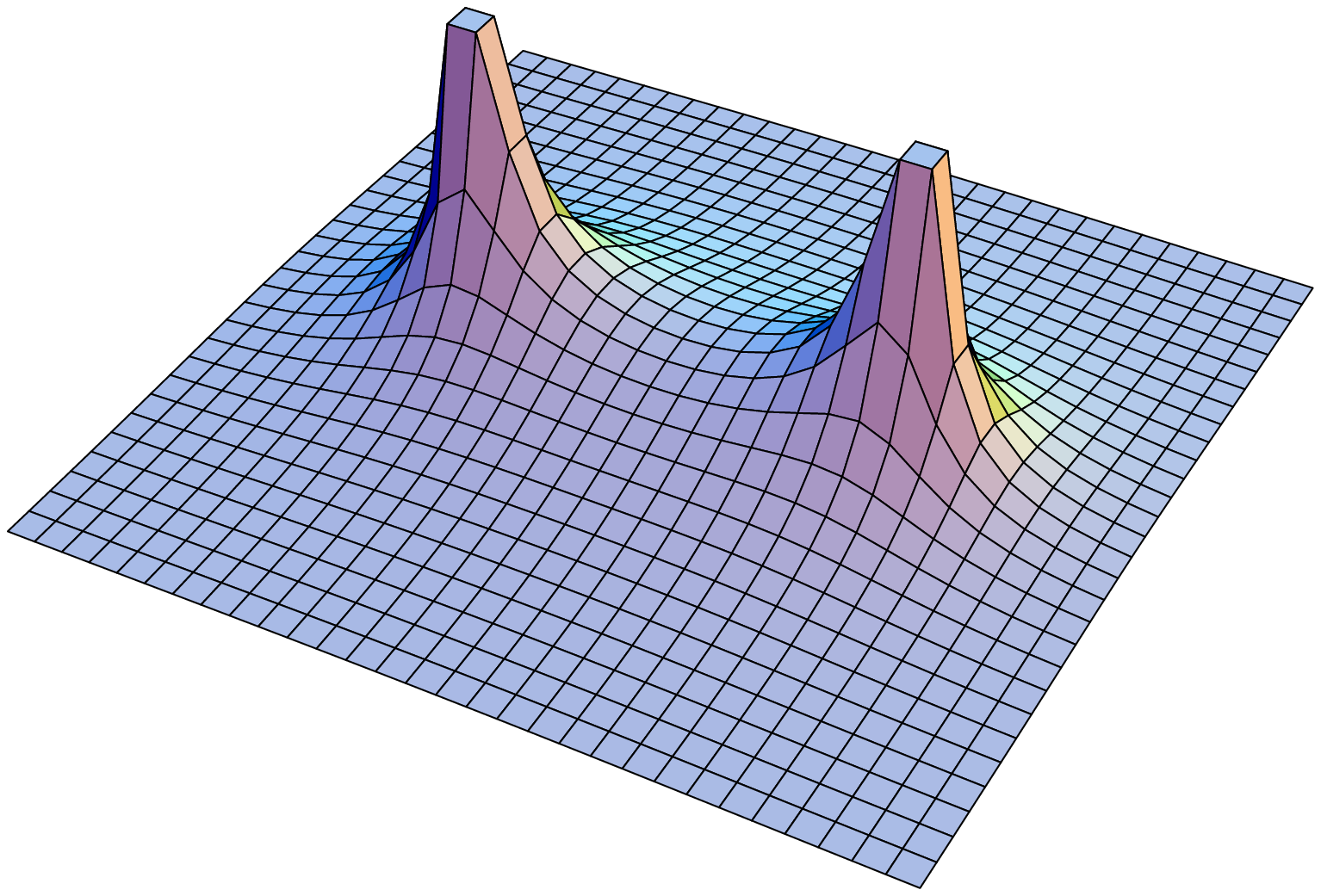}
\includegraphics{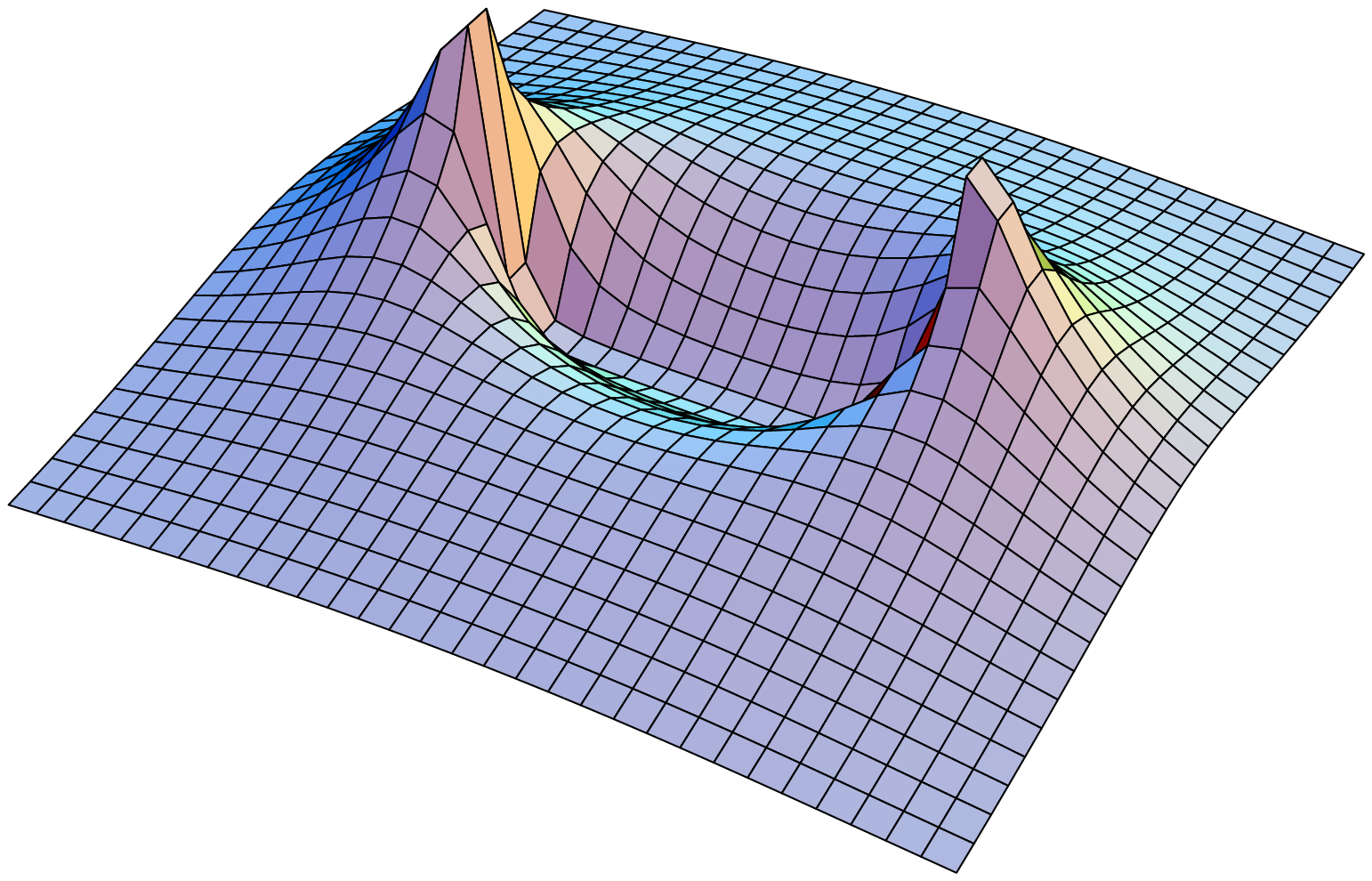}
\caption{The chiral zero-mode for the bipole of charge $k=1$ (left) and 
$k=2$ (right). Plotted is $|\Psi(\vec x)|^2$, the integral normalised to 
1, as a function of $\protect\sqrt{x^2+y^2}$ and $z$ on a linear scale 
for $s=.5$, cutoff at respectively $|\Psi|^2=5$ (left) and .5 (right).}
\label{fig:psi2}
\end{figure}

To find a normalisable zero-energy solution, we have to do something about the
asymptotic behaviour. A natural way to achieve this is to consider an abelian
BPS (self-dual) dipole, or {\bf bipole} for short. We now profit from having 
expressed the zero-mode in terms of the function $w$. The bipole field is 
generated by 
\beq
\log w(\vec x)=\log(|\vec x|+\vec x\cdot\hat n)-\log(|\vec x+s\hat n|+
\vec x\cdot\hat n+s)
\eeq
where the first term represents a monopole at $\vec x=\vec 0$ and the
second term, with the opposite sign, an anti-monopole at $\vec x=-s\hat n$.
It is convenient to express $w(\vec x)$ as
\beq
w(\vec x)=\frac{r_1+r_2-s}{r_1+r_2+s},\label{eq:wdip}
\eeq
where $\vec r_1=\vec x$ and $\vec r_2=\vec x+s\hat n$. This shows that the
Dirac strings of the monopole and anti-monopole partly cancel. All that we 
need to check is if $\Psi$, with this choice of $w$ is now normalisable. A 
simple computation shows that $|\Psi(\vec x)|^2=\Delta w(\vec x)=\frac{4s^2
}{r_1r_2(r_1+r_2+s)^2}$, as shown in Fig.~\ref{fig:psi2} (left). This is 
indeed integrable at the location of the two point charges and at infinity, 
$\int d^3x|\Psi(\vec x)|^2=\int d^3x \Delta w(\vec x)=4\pi s$, using that 
$w(\vec x)=1-s/|\vec x|+\cO(1/|\vec x|^3)$. 

Having found a normalisable zero-mode for one bipole, a generalisation to a 
collection of bipoles is obvious, by taking the product of $w(\vec x)$ for each
such bipole. This does not affect the property that $w(\vec x)$ is positive, 
vanishes along the Dirac strings, and its logarithm is harmonic elsewhere. 
However, the form of the zero-mode requires all factors $w$ to be formulated 
in terms of the same $\hat n$, which means all bipoles have to point in the 
{\em same} direction, i.e. the magnetic moments of all bipoles have to be 
uni-directional. It is not clear if this is just a limitation of our simple 
ansatz. For multi-bipoles, all separated much further than each of the 
individual bipole sizes ($s$), $|\Psi|^2$ will near each bipole be of the same 
form as for a single bipole. However, when two or more bipoles coincide, or 
equivalently when $g$ is bigger than the minimal Dirac value of $\frac{1}{2e}$,
$|\Psi|^2$ will be suppressed along the line segment connecting the two 
charges. To demonstrate this, we note that for $g=\frac{k}{2e}$, $w_k(\vec x)=
w^k(\vec x)$, with $w(\vec x)$ as given in Eq.~(\ref{eq:wdip}). Thus, using 
$w(\vec x)\Delta\log w(\vec x)=0$, we find 
\beq
|\Psi(\vec x)|^2=\Delta w(\vec x)^k=k^2w^{k-1}(\vec x)\Delta w(\vec x)
=\frac{4s^2k^2(r_1+r_2-s)^{k-1}}{r_1r_2(r_1+r_2+s)^{k+1}},
\eeq
which integrates to $4\pi ks$. The case for $k=2$ is shown in 
Fig.~\ref{fig:psi2} (right).

\section{The Caloron}

The context in which the bipole appears in a natural way is the caloron with 
non-trivial holonomy~\cite{PLB2}, in the infinite temperature limit. The 
periodic boundary conditions in the Euclidean time direction, relevant 
for these finite temperature instantons, allow for a non-trivial holonomy 
determined by the Polyakov loop, which approaches a constant value at spatial 
infinity,
\beq
\pl=\lim_{|\vec x|\rightarrow\infty}P(\vec x),\quad
P(\vec x)={\rm P}\exp(\int_0^\beta A_0(t,\vec x)dt).
\eeq
With $A_0$ playing the role of a Higgs field, a non-trivial value implies 
that an SU($n$) charge one caloron splits into $n$ constituent BPS monopoles, 
whose masses are determined by the eigenvalues of the Polyakov loop
\beq
\plo\equiv\exp(2\pi i\diag(\mu_1,\mu_2,\cdots,\mu_n)),\quad
\sum_{i=1}^n\mu_i=0.
\eeq
arranged to satisfy $\mu_1\leq\cdots\mu_n\leq\mu_{n+1}\equiv\mu_1+1$. The 
constituent masses $8\pi^2\nu_m$, with $\nu_m\equiv(\mu_{m+1}-\mu_m)/\beta$, 
add up to $8\pi^2/\beta$ such that the action equals that of a charge one 
instanton. The presence of these constituents is easily established from the
formula~\cite{PLB2,PLBN} 
\beqa
\Tr F_{\alpha\gamma}^{\,2}(x)=\partial_\alpha^2\partial_\gamma^2\log\psi(x),
\quad\psi(x)=\half\tr(\cA_n\cdots \cA_1)-\cos(2\pi t),\label{eq:action}\\
\cA_m\equiv\frac{1}{r_m}\left(\!\!\!\bea{cc}r_m\!\!&|\vec\rho_{m+1}|\\0
\!\!& r_{m+1}\eea\!\!\!\right)\left(\!\!\!\bea{cc}\cosh(2\pi\nu_m r_m)
\!\!&\sinh(2\pi\nu_m r_m)\\ \sinh(2\pi\nu_m r_m)\!\!&\cosh(2\pi\nu_m r_m)
\eea\!\!\!\right),\nonumber
\eeqa
where we introduced $r_m\equiv|\vec x-\vec y_m|$ and $\vec\rho_m\equiv\vec 
y_m-\vec y_{m-1}$ ($r_{n+1}\equiv r_1$ and $\vec y_{n+1}\equiv\vec y_1$), 
with $\vec y_m$ the location of the $m^{\rm th}$ constituent monopole with 
a mass $8\pi^2\nu_m$. 

The basic ingredient in the construction of caloron solutions is the Greens 
function $\hat f_x$ defined on the circle\footnote{For example $\Tr F_{\alpha
\gamma}^{\,2}(x)=-\partial_\alpha^2\partial_\gamma^2\log\det\hat f_x$, leading 
to the result of Eq.~(\ref{eq:action}).}, $z\in[0,\beta^{-1}]$, 
satisfying~\cite{PLB2,PLBN}
\beq
\left(\left(\frac{1}{2\pi i}\frac{d}{dz}-t\right)^2+r^2(\vec x;z)+\frac{1}{2\pi}
\sum_m\delta(z-\mu_m/\beta)|\vec\rho_m|\right)\hat f_x(z,z')=\delta(z\!-\!z'),
\eeq
where $r^2(\vec x;z)\!=\!r_m^2$ for $z\in[\mu_m/\beta,\mu_{m+1}/\beta]$. The 
variable $z$ can be introduced through Fourier transformation with respect to 
time, where the Fourier coefficients are related to the ADHM data~\cite{ADHM} 
of instantons, periodic up to a gauge rotation with $\pl$ (giving the solution 
in the so-called algebraic gauge). This is in one-to-one relation with the 
Nahm transformation~\cite{Nahm}. For $\mu_m/\beta\leq z'\leq z\leq\mu_{m+1}
/\beta$ ($\hat f_x(z',z)=\hat f_x^*(z,z')$ for $z<z'$) the explicit 
result~\cite{PLB2,MTP} for the Greens function can be expressed as
\beq
\hat f_x(z,z')=\frac{\pi e^{2\pi i t(z-z')}}{r_m\psi}\langle v_m(z')|\cA_{m\!
-\!1}\cdots \cA_1\cA_n\cdots \cA_m-e^{-2\pi it}|\sigma_2v_m(z)\rangle,
\label{eq:green}
\eeq
where the spinor $v_m(z)$ is defined by
\beq
v_m(z)=\pmatrix{\sinh[2\pi(z-\mu_m/\beta)r_m]\cr\cosh[2\pi(z-\mu_m/\beta)r_m]
\cr}.
\eeq
The chiral Dirac, or Weyl equation can be solved with the boundary condition 
$\Psi_z(t+\beta,\vec x)=\exp(2\pi i z\beta)\pl\Psi_z(t,\vec x)$ (in addition 
to the two component spinor index, there is now also a colour index). With 
$z=\half\beta^{-1}$ one obtains the finite temperature ``anti-periodic"
fermion zero-mode, and for $z=0$ the ``periodic" zero-mode.

To be specific, for the SU(2) caloron we have $\mu_2=-\mu_1\equiv\beta\omega$ 
and $|\vec\rho_1|=|\vec\rho_2|\equiv\pi\rho^2/\beta$ ($\rho$ is the instanton 
scale parameter). The gauge field and zero-mode can be expressed in terms of 
the functions $\phi^{-1}\equiv 1-\rho^2\hat f_x(\omega,\omega)/\beta$ and $\chi
\equiv\rho^2\hat f_x(\omega,-\omega)/\beta$. In the algebraic gauge, choosing 
$\pl=\plo$ and the constituents along the $z$-axis (by proper combinations of 
gauge and space rotations this can always be achieved)~\cite{PLB2}
\beq
A_\alpha=\frac{i}{2}\bar\eta^3_{\alpha\gamma}\tau_3\partial_\gamma\log\phi+
\frac{i}{2}\phi\myre\left((\bar\eta^1_{\alpha\gamma}-i\bar\eta^2_{\alpha\gamma})
(\tau_1+i\tau_2)\partial_\gamma\chi\right),\label{eq:ag}
\eeq
with $\bar\eta_{\alpha\gamma}^a\sigma_a\equiv\bar\sigma_{[\alpha}
\sigma_{\gamma]}$ the anti-selfdual 't Hooft tensor, and~\cite{MTCP}
\beq
\Psi_z(x)=\frac{\rho}{2\pi\beta}\sqrt{\phi(x)}\left(\pmatrix{\partial_2+i
\partial_1\cr\partial_0-i\partial_3\cr}\hat f_x(\omega,z),~\pmatrix{-\partial_0
-i\partial_3\cr\hphm\partial_2-i\partial_1\cr}\hat f_x(-\omega,z)\right).
\eeq
Particularly simple and valid for arbitrary SU($n$), is the expression for 
the density of the fermion zero-modes~\cite{MTP,MTCP}
\beq
|\Psi_z(x)|^2=-(4\pi^2\beta)^{-1}\partial_\alpha^2\hat f_x(z,z).
\label{eq:psi2}
\eeq

The limit $\beta\rightarrow0$ can be seen as a dimensional reduction and 
only the time independent field components are expected to survive in this 
high temperature limit. It is therefore more appropriate to consider the 
periodic gauge. For general $z$ this periodic gauge is obtained by applying 
the gauge transformation $g(t)\equiv e^{-tA_0^\infty}$, where $A_\alpha^\infty
\equiv2\pi i(\omega\tau_3-z)\delta_{\alpha0}$, such that $\tilde\Psi_z=g(t)
\Psi_z$ and the new gauge field $\tilde A$ are now periodic, with
\beq
\tilde A_\alpha=A_\alpha^\infty+\frac{i}{2}\bar\eta^3_{\alpha\gamma}\tau_3
\partial_\gamma\log\phi+\frac{i}{2}\phi\myre\left((\bar\eta^1_{\alpha\gamma}
-i\bar\eta^2_{\alpha\gamma})(\tau_1+i\tau_2)e^{-2\pi it\nu_1}\partial_\gamma
\chi\right),
\eeq
We find that~\cite{PLB2} $\lim_{\beta\rightarrow0}\chi(x)=0$ and $\lim_{\beta
\rightarrow0}\phi^{-1}(x)=w(\vec x)$ with $w(\vec x)$ as in Eq.~(\ref{eq:wdip}).
The resulting abelian gauge field splits into an isospin up and down component, 
{\em decoupled} in the Weyl equation. For $z=0$ this gives a mass of $\pi\nu_1$
to both isospin components, which contribute equally to the density, and the 
zero-mode is supported entirely at $\vec y_1$, whereas for $z=\half\beta^{-1}$ 
the mass is $\pi\nu_2$, but the zero-mode is now supported entirely at 
$\vec y_2$. For other values of $z$ the mass will depend on the isospin 
component, but as long as $z\neq\pm\omega$ the zero-mode remains localised to 
either of the two constituent locations, jumping from one to the other when 
$z$ crosses $\pm\omega$, where the zero-mode becomes delocalised, having 
support at both constituents simultaneously. Indeed, for $z=\omega$ 
\beq
\Psi_\omega(x)=\frac{\sqrt{\phi(x)}}{2\pi\rho}\left(\frac{1}{\phi(x)}
\pmatrix{\partial_2+i\partial_1\cr\partial_0-i\partial_3\cr}\log\phi(x),~
\pmatrix{-\partial_0-i\partial_3\cr\hphm\partial_2-i\partial_1\cr}\chi^*(x)
\right),
\eeq
whereas for $z=-\omega$, the same result follows after charge and isospin
conjugation, $\Psi^1_{-\omega}(x)=-\Psi^2_{\omega}(x)^\dagger\sigma_2$ and 
$\Psi^2_{-\omega}(x)=\Psi^1_{\omega}(x)^\dagger\sigma_2$ (with the isospin 
index made explicit). Since $\chi$ vanishes in the high temperature limit, 
the only surviving isospin component of the Weyl zero-mode is the one for 
which the asymptotic value of $A_0$ vanishes, and for which the zero-mode 
is time independent. This non-trivial isospin component agrees (up to an 
irrelevant factor $-2\pi i\rho$) with Eq.~(\ref{eq:azm}), for $w(\vec x)=
\phi^{-1}(x)$.

For the high temperature limit to be smooth and unambiguous, it was essential 
that $\phi$ be time independent, $\phi^{-1}\Delta\log\phi=0$, and that 
$\chi=0$. It is interesting to note that, imposing self-duality on
Eq.~(\ref{eq:ag}), leads to a natural generalisation at finite temperature 
\beqa
\phi^{-2}\partial_\alpha^2\log\phi+|(\partial_1-i\partial_2)
\chi|^2+|(\partial_0-i\partial_3)\chi|^2=0,\\
\phi^{-1}(\partial_1+i\partial_2)\phi^2(\partial_1-i\partial_2)\chi+
\phi^{-1}(\partial_0+i\partial_3)\phi^2(\partial_0-i\partial_3)\chi=0.
\nonumber
\eeqa
In principle, but not in practise, this can be used to define $\phi$ and 
$\chi$. Interestingly these equations also appear when formulating 
self-duality in the so-called R-gauge introduced by Yang~\cite{Yang}, 
after a suitable B\"acklund transformation~\cite{Back}. 

We will end with a few words on the case of SU($n>2$). In the high temperature
limit the zero-mode is again exponentially localised at one of the 
constituents~\cite{MTP}. This we can read off from Eq.~(\ref{eq:psi2}), using 
the explicit expression for $\hat f_x(z,z)$ given in Eq.~(\ref{eq:green}). 
When $z$ passes through $\mu_m/\beta$ the zero-mode jumps from one constituent 
location to the other, and only for these values of $z$ the zero-mode will 
delocalise, with the proviso that it will only ``see" {\em two} out of the 
$n$ constituents. In the periodic gauge, diagonalising $A_0$ at infinity by 
a constant gauge rotation, we have $A_0\rightarrow 2\pi i\diag(\mu_1/\beta-z,
\mu_2/\beta-z,\cdots,\mu_n/\beta-z)$. The only  non-vanishing colour component
of the fermion zero-mode is the one for which $\mu_j/\beta-z=0$. The resulting 
configuration is again that of the bipole in section 2.

\section{Discussion}
An interesting question is if there is some physical significance to the
zero-modes, like for chiral symmetry breaking in the effective description
of QCD in terms of monopoles obtained through abelian projection~\cite{AbPr}. 
But first we would like to better understand how to go beyond the case where 
the magnetic moments of the bipoles are no longer parallel. A natural setting 
in which to address this particular question would be through the study of 
charge $k>1$ calorons. It is not clear if in the high temperature limit a 
global abelian embedding for the case of non-parallel magnetic moments 
exists. Even when the magnetic moments are parallel, one would expect $k$ 
independent zero-energy solutions, of which we have only provided one. 

What led us to the results presented in this paper, was an attempt to solve 
the Weyl equation in the background of the abelian gauge field~\cite{PLBT} 
that is obtained from the Nahm transformation of an SU(2) charge one instanton 
on $T^3\times R$. This self-dual abelian gauge field is described by two 
bipoles on $T^3$, but unfortunately with {\em anti-parallel} orientations. 
Furthermore, two zero-modes are required in order for the (inverse) Nahm 
transformation to reconstruct the SU(2) instanton, for which $z$ is to be 
identified with the time. When the zero-modes are localised at the monopole 
singularities, it can be shown that the resulting SU(2) gauge field is 
abelian. This describes the asymptotically flat connections ($F=0$) of the 
instanton on $T^3\times R$, required for the integral over the action density 
to be finite, and equal to $8\pi^2$. Thus, to obtain genuine non-abelian 
behaviour the zero-mode has to become delocalised for certain values of $z$. 
Here the gauge field is no longer flat ($F\neq0$) and the density will have 
a maximum, in accordance with a general relationship between holonomies (with 
respect to each of the generating circles of the manifold) of a self-dual 
gauge field on the one hand and the constituent locations of the Nahm dual 
gauge field on the other hand. This dual relationship has been verified in a 
careful numerical study~\cite{Tdual}. Similar results have been 
seen~\cite{Ford} for instantons on $T^2\times R^2$.

It is the applications to the Nahm transformation on a torus that has been 
our prime motivation for studying this problem. Taken out of this context 
and the context of the caloron, our exact result for the chiral zero-mode 
is so simple we believe it is worthwhile to share it with the reader.

\section*{Acknowledgements}
I am grateful to \v Stefan Olejnik and Jeff Greensite for inviting me to a 
wonderfully well organised workshop in the beautiful setting of Slovakia's
``pearl", the Spis region with its High Tatra mountains. I also thank Maxim 
Chernodub for our attempts to find physical applications for these bipoles 
with parallel magnetic moments, Chris Ford for pointing me to Ref.~\cite{Back}, 
and them as well as Falk Bruckmann, Conor Hougton and Valya Zakharov for 
useful discussions. I am particularly grateful to Margarita Garc\'{\i}a 
P\'erez for her generous collaboration in attempting to apply the methods 
presented here to the Nahm transformation on $T^3\times R$ and for 
comments on a first draft of this paper.

\end{document}